\documentclass[aps,twocolumn,showpacs,superscriptaddress,groupedaddress]{revtex4-1}  
\usepackage{graphicx}  
\usepackage{dcolumn}   
\usepackage{bm}        
\usepackage{amssymb}   
\usepackage{amsmath}
\usepackage{epsfig}
\usepackage{color}
\usepackage{xspace}
\usepackage{cleveref}
\begin{document}

\title{Molecular alignment and filamentation: comparison between weak and strong field models}
\author{N. Berti$^{1,2}$}
\author{P. B\'ejot$^{1}$}\email{pierre.bejot@u-bourgogne.fr}
\author{J.-P. Wolf$^{2}$}
\author{O. Faucher$^{1}$}

\affiliation{$^{1}$ Laboratoire Interdisciplinaire CARNOT de Bourgogne, UMR 6303 CNRS-Universit\'e de Bourgogne, BP 47870, 21078 Dijon, France}
\affiliation{$^{2}$ Universit\'e de Gen\`eve, GAP-Biophotonics, Chemin de Pinchat 22, 1211 Geneva 4, Switzerland}

\begin{abstract}
The impact of nonadiabatic laser-induced molecular alignment on filamentation is numerically studied. Weak and strong field model of impulsive molecular alignment are compared in the context of nonlinear pulse propagation. It is shown that the widely used weak field model describing the refractive index modification induced by impulsive molecular alignment accurately reproduces the propagation dynamics providing that only a single pulse is involved during the experiment. On the contrary, it fails at reproducing the nonlinear propagation experienced by an intense laser pulse traveling in the wake of a second strong laser pulse. The discrepancy depends on the relative delay between the two pulses and is maximal for delays corresponding to half the rotational period of the molecule.
\end{abstract}

\maketitle

\section{Introduction}
Since its first experimental observation in gases in the mid-1990s \cite{Braun95}, laser filamentation, i.e., the nonlinear propagation of ultrashort intense laser pulses, has attracted extensive attentions due to its physical interest, as well as its important applications including few-cycle optical pulse generation, terahertz generation, supercontinuum generation, and remote sensing \cite{ChinReport,BergeReport,MysyReport,KasparianW08}. The main feature of filamentation is its ability to sustain very high intensities (around 50\,TW/cm$^2$) over very long distances in contrast with predictions of linear propagation theory. When exposed to such laser field intensities, atoms and molecules exhibit highly nonlinear dynamics leading to the observation of phenomena such as multiphoton and tunnel ionization, harmonic generation, and nonadiabatic molecular alignment. The last is a process that occurs when non-spherical molecules are exposed to a short and intense laser pulse \cite{Stapelfeldt2003}. The nonresonant interaction, driven by a pulse of duration much shorter than the classical rotational period, results in the production of postpulse transient molecular alignment revivals. The possibility of confining in space the rotational axes of a molecule, in the absence of the strong driving field, has been found particularly useful in various fields extending to high-harmonic generation and attophysics \cite{Itatani94_2005,Haessler6_2010}, molecular tomography \cite{Itatani432_2004,Meckel320_2008}, molecular-frame photoelectron angular distribution \cite{Bisgaard323_2009,Holmegaard_2010,Hansen106_2011},  control of molecular scattering \cite{Gershnabel104_2010}, and more generally study of direction-dependent interactions, to mention but a few.  Moreover, field-free molecular alignment results in a modification of the local optical refractive index, which has been used for controlling the propagation dynamics of weak beams \cite{Calegari2008,Cai2009_2,Wang2010} and filaments \cite{Wu2008,Wu2009,Peng2009,Cai2009,Calegari2009,Cai2010,Varma2012,Palastro2012}. Simulating laser pulse propagation over macroscopic distances in an aligned molecular medium is complicated by the need to include quantum-mechanical laser-molecule dynamics. While molecular alignment calculations are routinely performed with the time-dependent Schrod\"{\i}nger equation, its consideration in the context of two-dimensional laser propagation requires high numerical resources. Instead, the standard treatment of molecular alignment in propagation simulations consists of approximating the refractive index modification   with a perturbation model, in which the rotational populations of the system remain unchanged upon the laser field application. In this approximation,  the refractive index modification induced by molecular alignment is expressed as a convolution between the temporal pulse profile and the impulsive response of the molecule. This procedure provides a fast and efficient evaluation of the effect of molecular alignment that is convenient for  laser propagation simulations. While this approximation is valid for the weak field regime, the present work shows that it is no longer valid for  intensities encountered in a filament, i.e., a tens of TW/cm$^2$.

Nonlinear propagation simulations are compared using  the weak field  and the full quantum-mechanical treatment. Two cases can be distinguished. When a filament propagates through a thermal ensemble of molecules, the weak field model well reproduces the full quantum simulation. On the contrary, when a filament propagates through an ensemble of molecules previously aligned by a second laser filament, the weak field model fails at reproducing the propagation dynamics of the first filament. This is particularly the case when the delay between the two pulses matches the half rotational period of the molecule \cite{Lapert2009,Hoque2011}.

 The paper is divided as follows. The first section is devoted to the refractive index modification induced by molecular alignment using the strong and weak field model. In the second section, numerical simulations of filamentation in molecular gases are performed using a strong field modeling of molecular alignment. Their predictions are compared with those obtained with the ``standard" weak field model for the one- and two-pulse cases. It is shown that the weak field model well reproduces  the results obtained with the strong field model in the case of a single filament. On the contrary, it fails at reproducing pump-pump experiments in which a laser filament propagates in the wake of a second filament.

\section{Molecular alignment}
\subsection{Strong field model}

Laser-induced alignment of a molecular ensemble is described by solving the time-dependent Liouville equation
\begin{equation}
	\textrm{i}\dfrac{\partial}{\partial t} \varrho(t) = \big[ H_{\textrm{rot}}+H_{\textrm{int}}(t), \varrho(t) \big],
	\label{Liouville}
\end{equation}
where $\big[A,B\big]=AB-BA$ denotes the commutator operator, $\varrho(t)$ is the density matrix operator, and  $H_{\textrm{rot}} = B\textbf{J}^{2}-D\textbf{J}^{4}$  is the rotational Hamiltonian, with $\textbf{J}$ the angular momentum and $B$ ($D$)  the rotational (centrifugal distortion) constant. For a linear molecule and a linearly polarized radiation, the interaction Hamiltonien is defined by
\begin{equation}
	H_{\textrm{int}}=-\frac{1}{4} \varepsilon(t)^{2} \Delta\alpha \cos^{2}\theta,
	\label{Hint}
\end{equation}
with $\Delta\alpha$ representing the polarizability anisotropy of the molecule and $\theta$ the angle between the molecular and the laser polarization axis \cite{Rouzee}. The degree of molecular alignment with respect to the axis $z$ is evaluated through both a quantum and a thermal averaging  of the operator $\cos^2\theta$ as

\begin{equation}
\ \langle\langle \cos^{2}\theta \rangle\rangle =\textrm{Tr}\big(\varrho\cos^{2}\theta\big),
\end{equation}
where $\textrm{Tr}$ defines the trace operator.

The refractive index modification $\Delta n_{\textrm{r}}$ resulting from the molecular alignment is given by
\begin{equation}
	\Delta n_{\textrm{r}} = \frac{N \Delta \alpha}{2\epsilon_{0}} \ \langle\langle \cos^{2}\theta - 1/3 \rangle\rangle,	
\end{equation}
with $\epsilon_{0}$ the permittivity of vacuum and $N$ the number density.

\begin{figure}[h!]
	\centering
	\includegraphics[width=8.6cm]{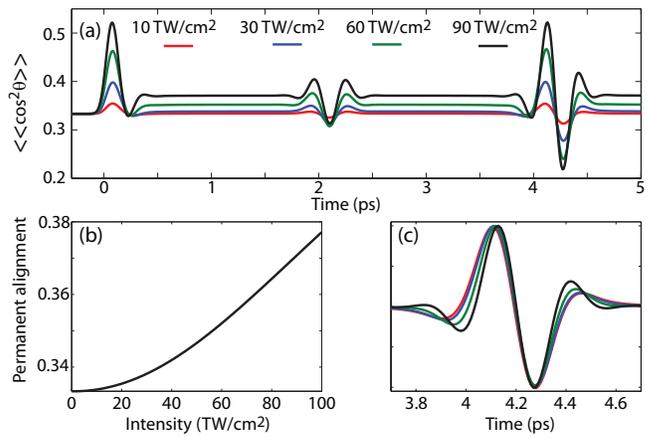}
	\caption{(Color online) (a) Temporal evolution of the molecular alignment parameter $\langle\langle\cos^2\theta\rangle\rangle$ induced in N$_2$ at 300 K by a 100 fs pulse with different peak intensities. (b) Permanent alignment as a function of intensity. (c) First revival for different peak intensities.}
	\label{ali_pump}
\end{figure}

As an example, Fig. \ref{ali_pump}(a) shows the alignment of N$_2$ calculated for different laser intensities. The molecular alignment can be split into two distinct components that differ by their temporal structures. The first one, called \textit{ alignment revival}, appears at each quarter of the rotational period $T_{\textrm{r}}$. It results from the periodic rephasing of the rotational wavepacket. The revival amplitude is proportional to the laser intensity in the weak field regime, i.e., below 60 TW/cm$^2$ for nitrogen at room temperature. One has to emphasize that whereas the shape of the revivals is constant in the weak field regime approximation,  the structural shape of the revival is no longer conserved  when the field becomes stronger, as shown in Fig. \ref{ali_pump}(c). The second component, called the \textit{permanent alignment}, comes from the populations redistribution among  rotational states. The interaction term defined in Eq.~\ref{Hint} only allows  $\Delta J=0, \pm2$ and $\Delta M=0$ transitions. Therefore, the result of the field action is to increase the total angular momentum $J$ while its projection $M$ remains constant. The permanent alignment is a direct consequence of the presence  in the wavepacket of  aligned rotational state  with $J\gg M$. The permanent component becomes significant for intense electric field and  scales approximatively as the square of the laser intensity, as shown in Fig. \ref{ali_pump}(b).

\subsection{Weak field model}

Calculation of two-dimensional  propagation in  laser-aligned molecules  requires high numerical resources. To circumvent this problem,  a standard approach consists  in approximating the refractive index change induced by the molecular alignment with perturbation theory.

\subsubsection{Perturbation theory}

\begin{figure}[t!]
	\centering
		\includegraphics[width=8.6cm]{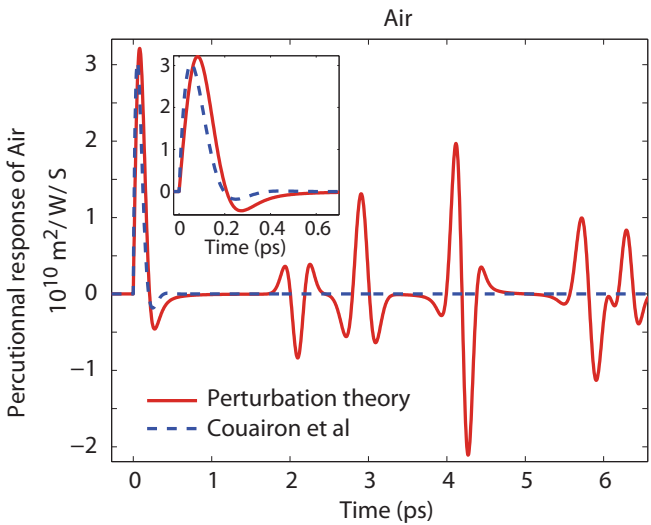}
		\caption{(Color online) Impulse response of air at 300 K evaluated with the perturbation theory (red solid line) and  the three step model (blue dashed line).}
		\label{impulse_response}
	\end{figure}

The perturbative expression of the nonlinear refractive index change induced by molecular alignment is evaluated as \cite{Morgen,Lin}
\begin{eqnarray}
&\Delta n_\textrm{r}(t)=\\ \nonumber
&\frac{N\Delta\alpha^2}{15\hbar\epsilon^2_0cn}&\sum_J{K_J\textrm{Im}\left(e^{i\omega_Jt}\int_{-\infty}^t{I(t')e^{-i\omega_Jt'}dt'}\right)},
\end{eqnarray}
where $I(t)$ is the laser intensity and  $K_J$ a factor defined by
\begin{equation}
K_J=g_J\left(\rho_{J+2}-\rho_{J}\right)\frac{(J+1)(J+2)}{2J+3},
\end{equation}
with $\rho_J$  the initial population of  level $J$, $g_J$  the nuclear spin degeneracy factor, c the speed of  light, $n$  the linear refractive index, and $\omega_J$ the Raman angular frequency between $J$ and $J+2$ levels. By defining the impulse response of the molecules $R(\tau)$ as
\begin{equation}
R(\tau)=\frac{N\Delta\alpha^2}{15\hbar\epsilon^2_0cn}\textrm{Hea}(\tau)\sum_J{K_J\sin{\omega_J\tau}},
\label{Percu_formula}
\end{equation}
where $\textrm{Hea}(\tau)$ is the Heaviside function, the change of refractive index can be obtained through a convolution with $I(t)$:
\begin{equation}
\Delta n_\textrm{r}=R(t)\ast I(t).
\label{Indice_convol}
\end{equation}
The knowledge of the impulse response $R$ then allows to evaluate the nonlinear refractive index change in the weak field limit. Figure \ref{impulse_response} displays the impulse response of air at ambient conditions calculated within the weak field framework. It has been obtained by adding the relative contributions of nitrogen and oxygen molecules.

\subsubsection{Three-level model}
\begin{figure}[t!]
	\centering
		\includegraphics[width=7cm]{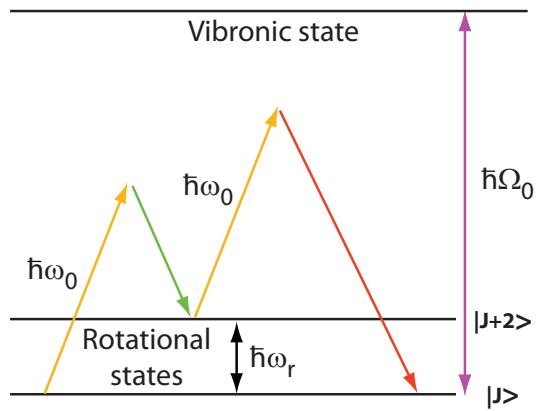}
		\caption{(Color online) Three-level system used for the estimation of the rotational contribution to the nonlinear refractive index in the weak field limit. The two populated rotational states $J$ and $J+2$ are coupled through  Raman-like transitions.}
		\label{3-level}
	\end{figure}

The nonlinear refractive index can also be estimated by describing  molecular alignment  as a three-level nonresonant process \cite{Penano}, as depicted in Fig.~\ref{3-level}.
$\Delta n_\textrm{r}$ is then evaluated through the differential equation
\begin{equation}
	\dfrac{\partial^{2} \Delta n_{\textrm{r}} }{\partial t^{2}} + 2\Gamma \dfrac{\partial \Delta n_{\textrm{r}}}{\partial t} + \Omega^{2} \Delta n_{\textrm{r}}  = \frac{\omega_r\mu^{2}}{\hbar^2\Omega_0^2}|\varepsilon|^{2},
	\label{Dn1}
\end{equation}
where $\mu$ is the transition dipole  moment matrix element related to the transition energy $\hbar\Omega_0$,  $\omega_0$ is the angular frequency of the field $\varepsilon$, and  $\Omega^{2} = \omega_{r}^{2} + \Gamma^{2}$ with $\omega_{r}$ the Raman angular frequency and  $\Gamma$ a  phenomenological  dephasing rate. In the frequency domain, the nonlinear refractive index $\widetilde{\Delta n}$ induced by the rotation is
\begin{eqnarray}	
\widetilde{\Delta n_{\textrm{r}} } &=& \frac{\omega_r\mu^{2}}{\hbar^2\Omega_0^2}\frac{1}{\Omega^{2}-\omega^{2}+ \mathrm{i}2\Gamma\omega} \widetilde{|\varepsilon|^{2}} \\
&=& \chi(\omega) \widetilde{|\varepsilon|^{2}}.
\end{eqnarray}	
In the temporal domain, the nonlinear refractive index is therefore described as in Eq. \ref{Indice_convol}, where the impulse response $R(t)$ is written as
\begin{eqnarray}
R(t)&=&\mathcal{R}_0\textrm{Hea}(t)\textrm{exp}(-\Gamma t)\textrm{sin}(\Omega t).
\label{Rt_Mysy}
\end{eqnarray}
The blue dashed curve in Fig. \ref{impulse_response} shows the impulse response of air used in \cite{MysyReport} and calculated according to Eq. \ref{Rt_Mysy}. It is in qualitative agreement up to 500 fs with the response function calculated with Eq. \ref{Percu_formula}. Even if this simplified model does not describe the periodic molecular alignment revivals, it remains so far the most widely used in filamentation simulations.

\subsection{Molecular alignment: weak vs strong field}
\begin{figure}[t!]
	\centering
		\includegraphics[width=8.6cm]{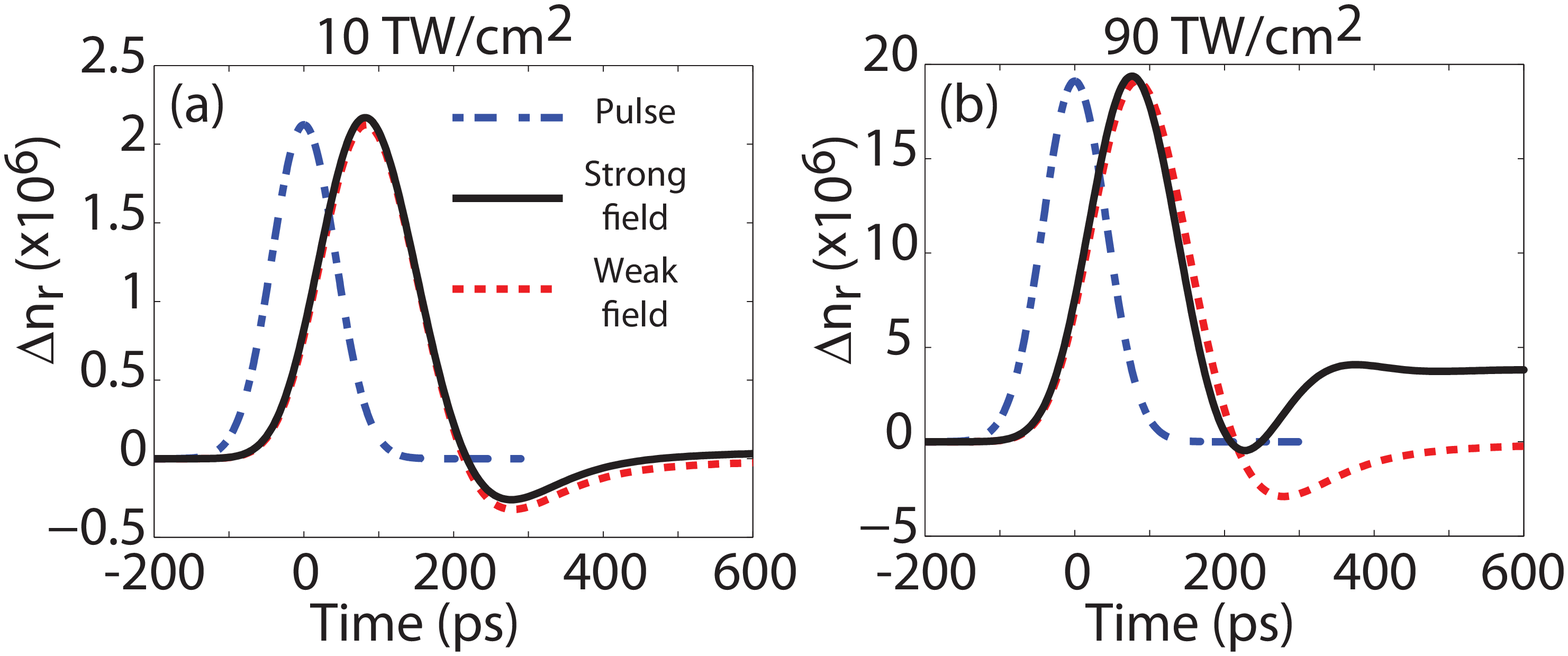}
		\caption{(Color online) Nonlinear refractive index change induced by molecular alignment as a function of time evaluated  by the weak (black solid lines) and the strong (red dashed lines) field theory for a 100 fs pulse. The peak intensity is  (a) 10 TW/cm$^{2}$  and  (b) 90 TW/cm$^{2}$.}
		\label{Error_ali_pump}
\end{figure}

The alignment of N$_2$ at 300 K induced by a single 100 fs gaussian pulse is first investigated. As shown in Fig. \ref{Error_ali_pump}(a), the nonlinear refractive index calculated in the weak field regime (Eq. \ref{Indice_convol}) is in good agreement with the strong field calculation. For a larger intensity, as shown in  Fig. \ref{Error_ali_pump}(b), the discrepancy between the two models is due to the permanent alignment effect that is not taken into account in the perturbation theory. However, this contribution  mainly affects the index   after the field extinction, so that it only marginally impacts the pulse propagation dynamics. The weak field theory is thus  well suited for nonlinear propagation simulations in the case of a single pulse.

\begin{figure}[h!]
	\centering
		\includegraphics[width=8cm]{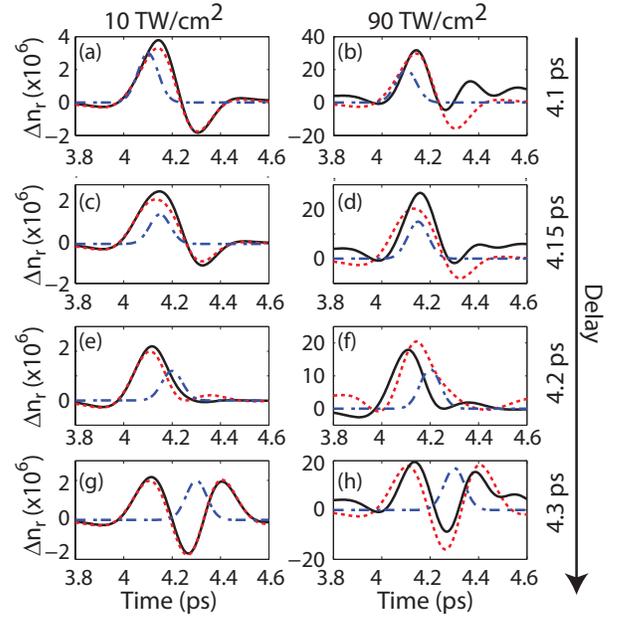}
		\caption{(Color online) Temporal evolution of the nonlinear refractive index  induced by molecular alignment evaluated  by the weak (red dashed lines) and strong (black solid lines) field theory for different relative delays between the two pulses. The two pulses share the same peak intensity: 10 TW/cm$^2$ (left) and 90 TW/cm$^2$ (right). The dotted-dashed blue lines depict the probe pulse temporal intensity distribution.}
		\label{Error_ali_probe}
	\end{figure}

In the case of two delayed laser pulses, the molecular alignment can exhibit a complex dynamics. As already shown in Fig. \ref{ali_pump}(c), the weak field approximation fails in  reproducing both  the deformation of the quantum revivals and the permanent alignment, which can have a large impact on a two-pulse experiment. Figure \ref{Error_ali_probe} depicts the nonlinear refractive index experienced by the second pulse evaluated either within the weak or  the strong field framework. Calculations have been performed for 100 fs (FWHM) collinearly polarized pulses as a function of the peak intensity and time delay. At low intensity (10 TW/cm$^{2}$), the permanent alignment is weak and the discrepancy between the two models remains marginal. This confirms that the perturbation framework is justified below this peak intensity value. It is clear that it does  not apply  at large intensity (90 TW/cm$^{2}$), where the discrepancy between the two models depends on the relative delay between the two pulses. As it is shown, the deviation of the weak field model prediction is maximal at a delay corresponding to half the rotational period ($\simeq$ 4.2 ps).
	
\section{Impact on the filamentation model}
	\subsection{Theory}

Assuming a cylindrical symmetry around the propagation axis, the equation driving the propagation of a linearly polarized electric field envelope $\varepsilon$ reads in the reciprocal space \cite{Kolesik2004}
\begin{multline}
\partial_{z}\widetilde{\varepsilon}= \mathrm{i}(k_{z}-\dfrac{\omega}{v_\mathrm{g}}) \widetilde{\varepsilon} + \dfrac{\omega}{c^{2}k_{z}} \bigg[\textrm{i}\omega \Big(
n_{2} \widetilde{|\varepsilon|^{2}\varepsilon} + \Delta \widetilde{n_{\textrm{r}}\varepsilon} \Big) \\
-\dfrac{e^{2}}{2\epsilon_{0} m_\mathrm{e}} \zeta(\omega)\ \widetilde{\rho \ \varepsilon} \bigg] - \widetilde{L[\varepsilon]},
\label{eq_UPPE}
\end{multline}
with $v_\mathrm{g}$ the group velocity, $e$ ($m_{\textrm{e}}$) the charge (mass) of the electron, $k_{z}=\sqrt{\textrm{k}^{2}(\omega)-\textrm{k}_{\bot}^{2}}$ with $k(\omega)$ the wave vector and $k_{\bot}$ its transversal component, $\zeta(\omega) = (\nu_{\mathrm{en}}+ \mathrm{i}\omega) / (\nu_{\mathrm{en}}^{2}+\omega^{2})$, where  $\nu_{\mathrm{en}}$ is the collision frequency between free electrons and neutrals atoms. The free-electron density $\rho$ follows
\begin{equation}
	\partial_{t} \rho = \textrm{W}(|\varepsilon|^{2})(N - \rho) + \dfrac{\sigma}{U_{i}}|\varepsilon|^{2} - g(\rho),
\end{equation}
where $\textrm{W}(|\varepsilon|^{2})$ describes the probability of ionization calculated with the PPT formula \cite{PPT}, $\sigma$ is the inverse Bremsstrahlung cross-section, $U_{\textrm{i}}$ is the ionization potential, $N$ is the numerical density of  molecules, and $g$ is the recombination function. The ionization rate was assumed to be insensitive to the molecular alignment \cite{Zhao2003,Pavicic2007}. The last term in equation~(\ref{eq_UPPE}) accounts for ionization-induced losses, and is calculated as
\begin{equation}
L[\varepsilon]=\frac{U_{\mathrm{i}}W(|\varepsilon|^{2})}{2|\varepsilon|^{2}}(N-\rho)\varepsilon.
\end{equation}
The change of refractive index induced by the molecular alignment  is evaluated  by using  both the weak (Eqs. \ref{Percu_formula}-\ref{Indice_convol}) and   the strong field theory.
In the last,  Eq.~\ref{Liouville} was solved  on spatial grid points where the pump fluence was higher than 0.05 J/cm$^{2}$, otherwise the weak field model was used in order to reduce the computational time. Note that the convergence of the calculations was carefully checked by performing the strong field calculations also for pump fluence higher than 0.02 J/cm$^2$.

\subsection{Results and discussion}
	\subsubsection{Single pulse case}
\begin{figure}[h!]
	\centerline{
		\includegraphics[width=8.6cm]{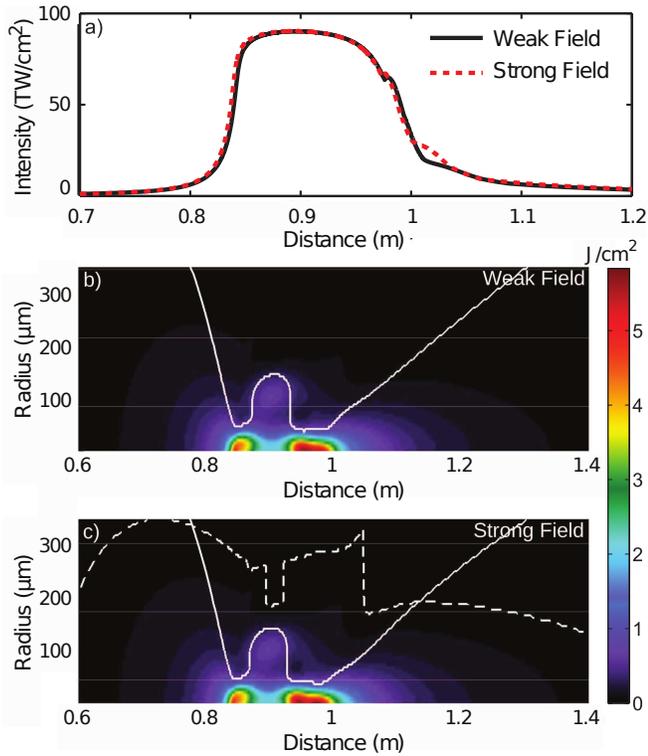}
			   }	
	\caption{(Color online) (a) Peak intensity  as a function of the propagation distance of a single 100 fs pulse using  the weak (black solid line) and  strong (red dashed line) field model. Fluence distribution as a function of the propagation distance calculated with the  (b) weak  and (c)  strong field model. The white solid lines represent the radius of the beam and the white dashed line depicts the limit of strong field calculations (see text).}
	\label{ProbeAlone}
\end{figure}
A 800 nm single pulse experiencing filamentation is considered. The 0.6 mJ 100 fs gaussian pulse is focused with a 1 m focal length and propagates through gaseous nitrogen (4 bar, 300 K). The evolution of the on-axis intensity along the propagation is depicted in Fig. \ref{ProbeAlone}~(a). The simulations are performed  according to  the strong and weak field model. In both cases, the nonlinear propagation model predicts a strong clamping of the intensity (around 90 TW/cm$^2$) induced by the dynamic equilibrium between the  focusing and defocusing contributions to the refractive index. A slight deviation of the weak field model from the strong field prediction can be noticed at the falling part of the curves. As mentioned before, the strong field calculations were limited to a spatial region where the pump fluence is larger than 0.05 J/cm$^2$. This limit is represented by the white dashed line in Fig. \ref{ProbeAlone}~(c). As shown in Figs. \ref{ProbeAlone}~(b) and (c), the weak field model also reproduces quite accurately the fluence distribution all along the propagation axis. These results confirm that the  weak field model is well suited for single pulse simulations.
\begin{figure}[t!]
	\centerline{
		\includegraphics[width=8.6cm]{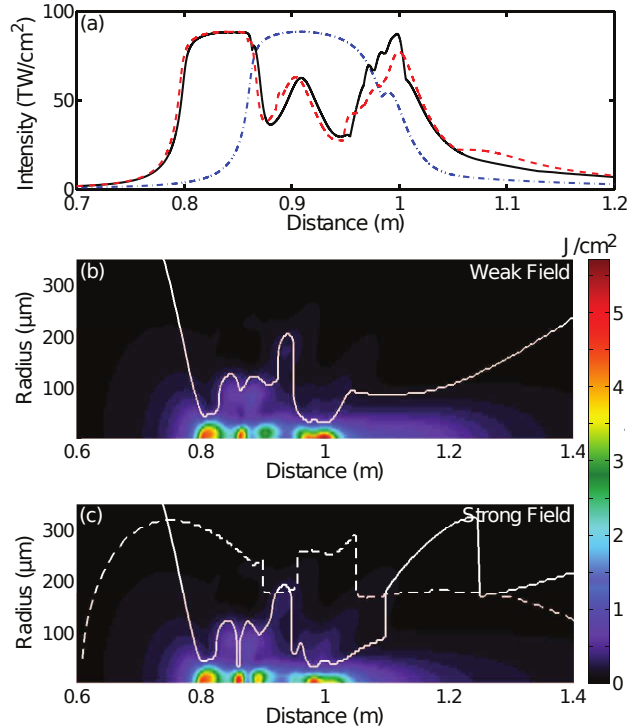}
			   }	
	\caption{(Color online) (a) Peak intensity along the propagation distance of the pump (blue dash-dotted line) and  probe  evaluated with  the weak (black solid line) and  strong (red dashed line) field theory. Panels (b) and (c) depict the probe fluence distribution along the propagation distance in the weak  and strong field regime, respectively. The white solid lines represent the radius of the beam and the white dashed line depicts the limit of strong field calculations. The pump-probe delay is set to $\tau$=4.1 ps. }
	\label{4ps1Flu}
\end{figure}
\subsubsection{Double pulse case}
\begin{figure}[t++!]
	\centerline{
		\includegraphics[width=8.6cm]{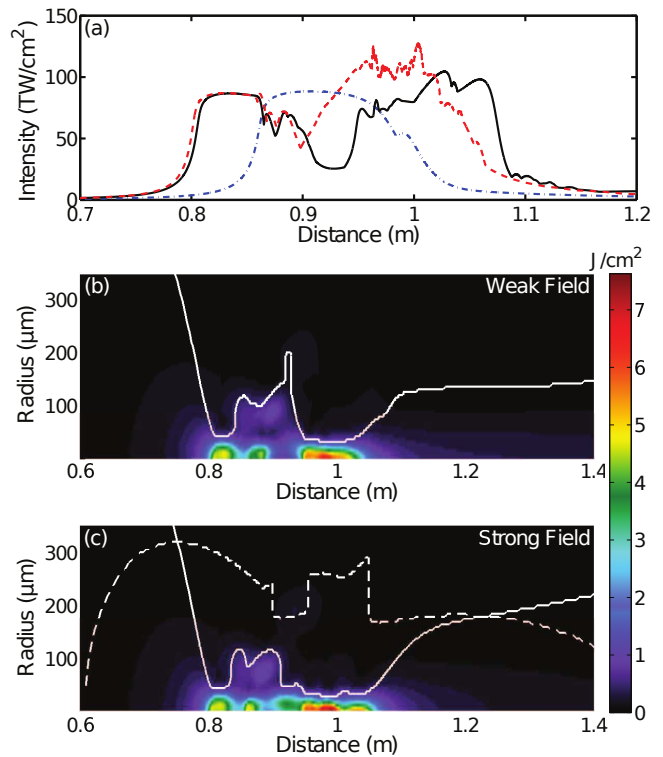}
			   }	
	\caption{(Color online) Same as Fig. \ref{4ps1Flu} except for the pump-probe delay set to  $\tau$=4.15 ps.}
	\label{4ps15Flu}
\end{figure}

\begin{figure}[h!]
	\centerline{
		\includegraphics[width=8.6cm]{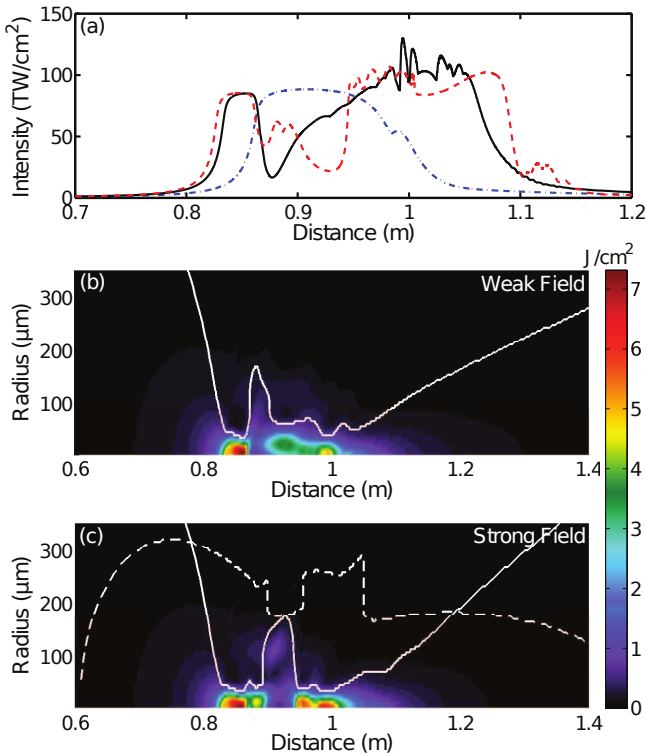}
			   }	
	\caption{(Color online) Same as Fig. \ref{4ps1Flu} except for the pump-probe delay set to  $\tau$=4.2 ps.}
	\label{4ps2Flu}
\end{figure}

We now consider  the filamentation dynamics of a pulse (denoted hereafter as a probe) traveling through a  medium previously aligned by a second filament (denoted hereafter as a pump). 
The pump (probe) energy is 0.5 (0.6) mJ. Different relative delays between the pump and probe beam are considered, so that the pump-induced molecular alignment  acts  as  either a focusing or a defocusing lens. Note that the probe propagation dynamics without the pump correspond to the case depicted in Fig.~\ref{ProbeAlone}. Depending on the relative delay between the pump and probe beam, the position and length of the probe filament is differently affected. This depends on the sign of the nonlinear refractive index experienced by the probe, which  is  either positive, when the molecules are  aligned, or negative, when they are  delocalized around the field axis.    For instance,  the data presented   in Fig. \ref{4ps1Flu} correspond to a relative delay of $\tau=4.1$ ps for which the pump-induced molecular alignment acts as a focusing lens ($\Delta n_\textrm{r}>$0). In comparison to the single pulse experiment, the probe filament sustains high intensities on a longer distance and its onset  is shifted backward.
\begin{figure}[t!]
	\centerline{
		\includegraphics[width=8.6cm]{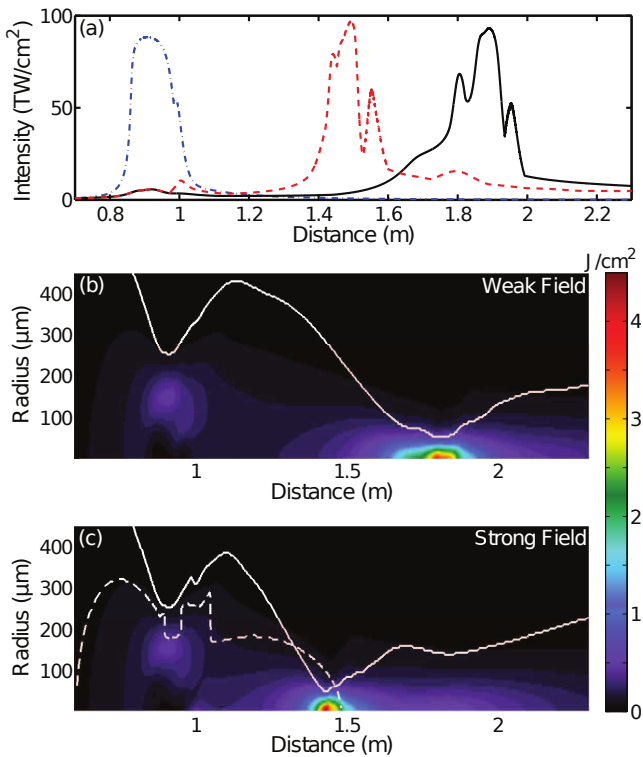}
			   }	
	\caption{(Color online) Same as Fig. \ref{4ps1Flu} except for the pump-probe delay set to  $\tau$=4.3 ps.}
	\label{4ps3Flu}
\end{figure}
Note that the weak and strong field models predict very similar results in this case. This is consistent  with the fact that the weak field model accurately reproduces the nonlinear refractive index calculated at $\tau$=4.1 ps [see Figs. \ref{Error_ali_probe} (a) and  (b)]. Figure \ref{4ps3Flu} presents the same results but  calculated for $\tau$=4.3 ps. In that case, the pump-induced molecular alignment acts as a defocusing lens ($\Delta n_\textrm{r}<$0),  inducing a strong shift of the probe filament position. Moreover, Figs. \ref{4ps3Flu} (b) and (c) show that the predictions of both models completely differs at the quantitative level. For instance, the weak field model predicts that the probe filament starts about 40 cm before the position predicted by the strong field model. This discrepancy is due to the fact that the weak field model fails at evaluating the nonlinear refractive index induced by molecular alignment at this particular delay [see Figs. \ref{Error_ali_probe} (g) and (h)]. Similar  disagreements are also reported at other delays lying around the half rotational period, as shown, for instance, in Figs. \ref{4ps15Flu} and \ref{4ps2Flu}.
These results  highlight the limit of the weak field model and show that a careful attention must be paid when evaluating the effect of molecular alignment on  two-pulse  experiments.

\section{Conclusion}
In this paper, the impact of molecular alignment on the propagation dynamics of a filament produced  in a one- and two-pulse configuration is studied. For single pulse, the weak field model  approximation provides a fair description of  the refractive index change induced by molecular alignment. For  double pulse,  i.e., when  a probe-filament propagates trough a medium previously aligned by a pump-filament,  it is shown that the propagation dynamics of the latter is strongly influenced by the former  that acts either as a focusing or a defocusing lens, depending on the relative delay between the two pulses. At some specific delays, in particular when they lie around the half rotational period  of the molecule, the weak field model is unable to  capture the dynamics of the probe filament. This work therefore  highlights the limit of weak field calculations in filamentation simulations.

\section*{Acknowledgment}
This work was supported by the Conseil R\'egional de Bourgogne (PARI program), the CNRS, the French National Research Agency (ANR) through the CoConicS program (contract No ANR-13-BS08-0013) and the Labex ACTION program (contract No ANR-11-LABX-0001-01). P.B. thanks the CRI-CCUB for CPU loan on its multiprocessor server. J.-P. W. acknowledges financial support from the European Research Council Advanced Grant ``Filatmo". The authors gratefully acknowledge E. Hertz for very fruitful discussions. The assistance of M. Moret was highly appreciated.

\end{document}